\documentclass[prb,12pt]{revtex4}

\usepackage[english]{babel}
\usepackage[intlimits]{amsmath}
\usepackage[dvips]{graphicx}

\begin{document}

\begin{titlepage}

\title{\bf {A Simple Kinetic Model Describes the Processivity of Myosin-V}}
\author{Anatoly B. Kolomeisky$^{*}$ and Michael E. Fisher$^{\dagger}$}
\affiliation{$^{*}$ Department of Chemistry, Rice University, Houston, TX 77005-1892\\
$^{\dagger}$Institute for Physical Science and Technology, University of Maryland,\\
 College Park, \ MD 20742}

\begin{abstract} 

Myosin-V is a motor    protein responsible  for organelle and vesicle  transport in cells. Recent single-molecule   experiments have shown that it is an  efficient processive  motor  that   walks   along actin filaments  taking steps of mean size close to  36 nm. A theoretical study of myosin-V motility is presented following an approach  used successfully to analyze the dynamics  of conventional kinesin but also taking some account of step-size variations.  Much of the  present  experimental data  for myosin-V   can be well described by a  two-state chemical kinetic model with three load-dependent rates. In addition, the analysis predicts the variation of the mean velocity and of the randomness --- a quantitative measure of the stochastic deviations from uniform, constant-speed motion --- with ATP concentration under both resisting and assisting loads, and  indicates a  {\it sub}step of size  $d_{0} \simeq$ 13-14 nm (from the ATP-binding site)  that appears to  accord  with  independent observations.

\end{abstract}

\maketitle

\end{titlepage}

\section*{INTRODUCTION}

Various classes of enzymes,  usually termed motor proteins, play important roles in biological processes such as cellular transport, cell division, muscle function and genetic transcription (see, e.g., Lodish et al., 1995). What we may term {\it translocatory} motor proteins (in contrast to rotary motor proteins) are epitomized by kinesins, dyneins, myosins, and DNA and RNA polymerases that move under loads along polar linear tracks such as microtubules, actin filaments, and double-stranded DNA, the motion being fueled by the hydrolysis of ATP or related reactions.

Motor proteins may work collectively in large groups, like myosin in muscles, or they may operate individually as do most microtubule-based kinesin and dynein molecules (Leibler and Huse, 1993; Howard, 2001). Those motor proteins that function collectively are typically {\it non}processive, i.e., they make at most one mechanical step along their tracks during a catalytic cycle before detaching from the track. On the other hand, individual motors that move vesicles over long distances (up to several microns) need to stay bound to their tracks over many catalytic cycles: such motors are {\it processive}. For example, conventional kinesin motors can walk along microtubules taking a hundred or more 8.2 nm steps before dissociating (Howard {\it et al.}, 1989; Block {\it et al.}, 1990; Vale {\it et al.}, 1996).

Recently, single-molecule experiments by Mehta {\it et al.} (1999; Mehta, 2001), Rief {\it et al.} (2000), Sakamoto {\it et al.} (2000), Rock {\it et al.} (2001),  Veigel {\it et al.} (2002) and Nishikawa {\it et al.} (2002) have demonstrated that myosin-V and myosin-VI, in contrast to the behavior of other members of the myosin superfamily (Howard, 2001), are also efficient processive molecular motors. Here we will focus on the dynamics of myosin-V.

Myosin-V is a dimeric, two-headed molecule that in the presence of actin readily hydrolyzes ATP to produce ADP and P$_{\mbox{i}}$ (Mehta, 2001). Kinetic experiments in bulk solution (De La Cruz {\it et al.}, 1999, 2000; Mehta, 2001) have demonstrated   that release of ADP is the rate-limiting step in the actin-activated ATPase cycle. Under conditions of limiting ATP the kinetically prevalent state appears to have  both head domains bound to the actin filament as captured in electron micrographs by Walker {\it et al.}, (2000); but more generally, see the discussions in Mehta (2001) and De La Cruz {\it et al.} (2001). 

Optical traps equipped with electronic feedback mechanisms have provided valuable information regarding the dynamics of individual myosin-V molecules under low load (see Mehta, 2001). The experiments allow one to monitor the displacement, $x(t)$, of a single molecule as a function of the time $t$ under different concentrations of ATP, etc., while maintaining a steady external load, $F$, which opposes the directed motion of the motor.

The principle experimental findings can be summarized as follows:  \  (i) Myosin-V moves along actin filaments towards the plus or barbed end taking large steps of size averaging 35-38 nm (Mehta {\it et al.}, 1999) approximating the 37 nm pseudo-repeat of the actin filament (Bray, 2001); \ (ii) The stepping dynamics depends strongly on the ATP concentration: thus, the mean dwell time $\tau(F, \mbox{[ATP]})$ observed between successive steps (preceding a forward step) at low [ATP] (=1 $\mu$M) hardly varies with the external load, while under saturating conditions ([ATP] $\geq$ 2 mM) the mean dwell time grows rapidly as $F$ approaches the stall force, $F_{S}=3.0 \pm 0.3$ pN (at which, on average, the motor just fails to progress); \ (iii) The overall stepping rate or mean velocity
\begin{equation}
V(F, \mbox{[ATP]}) \approx d \langle x(t) \rangle /dt
\end{equation}
follows a Michaelis-Menten form in that it is proportional to [ATP] at low concentrations but becomes independent of [ATP] under saturating conditions; \ (iv) Tight coupling between chemical and mechanical cycles is valid, with one ATP molecule hydrolyzed per individual myosin-V forward step along an actin filament; but \ (v) in contrast to the dynamics of conventional kinesin (Coppin {\it et al.}, 1997; Visscher {\it et al.}, 1999), myosin-V under load not infrequently exhibits sequences of two or three reverse or backward steps; and, finally, \ (vi) the addition of ADP to the {\it in vitro} solution significantly reduces the turnover rate of ATP (as is to be expected); moreover, the inhibitory effect of ADP scales with the concentration of ATP (Rief {\it et al.}, 2000) (and even when the mean stepping rate is reduced two-fold the distribution of dwell periods is unaltered).

The growing quantity of information concerning myosin-V has naturally stimulated theoretical discussions of the dynamics. Several models have been proposed and are reviewed by Mehta (2001). In particular, in order to provide an explanation of the observed load-dependence of the processivity, the mean dwell time at temperature $T$ has been modeled phenomenologically [following a proposal of Wang {\it et al.} (1998)] as a sum of two terms, namely
\begin{equation}
\tau(F) = \tau_{1}+\tau_{2} \exp(Fd'/k_{B}T),
\end{equation}
corresponding, respectively, to putative force-independent and force dependent transitions. It is natural to expect here that $d'$ corresponds to the observed step size $d \simeq 36$ nm (Fisher and Kolomeisky, 1999; Fisher and Kolomeisky, 2001; Hille, 2001). However, fitting the experimental data of Mehta {\it et al.} (1999), which is displayed in \mbox{Fig. 2} below, necessitates an (effective) step size $d'$ of 10-15 nm, which is only 30-40 $\%$ of the actual step size. This discrepancy is rationalized by asserting that $d'$ is some ``characteristic distance over which load affects the catalysis rate.'' Furthermore, this approach fails  to account clearly for the observed stalling of the motors at $F_{S} \simeq 3.0$ pN. Clearly, a more soundly based quantitative theory for processivity of myosin-V seems called for in order to satisfactorily describe the currently available data and to provide testable predictions. The present article aims to meet these requirements.

We present a theoretical analysis of the dynamics of myosin-V using simple, discrete-state stochastic models which have recently been developed and analyzed in detail by Kolomeisky and Widom (1998), Fisher and Kolomeisky (1999a, b; 2001; 2002) and Kolomeisky and Fisher (2000a, b); for brevity these articles will be referenced below as {\bf FK'01}, {\bf KF'00a}, etc. This approach has been used successfully in {\bf FK'01} to analyze the extensive experimental data on the dynamics of single conventional kinesin molecules moving {\it in vitro} along microtubules obtained by Visscher {\it et al.} (1999) and Schnitzer {\it et al.} (2000). We will demonstrate that most of the currently available experimental data on the processivity of myosin-V can be well accounted for by the simplest ($N=2$)-state model embodying a theoretical picture in satisfactory accord with other kinetic and structural experiments. Our treatment also provides specific predictions for as yet unexplored features of myosin-V dynamics that can be tested experimentally and should uncover further details of the stepping mechanism.

\section*{Theoretical Approach}

For completeness we first outline briefly the class of stochastic models used in our analysis and the {\it explicit} analytical results available for them. In the simplest periodic sequential kinetic model, illustrated schematically in \mbox{Fig. 1}, the protein motor is viewed as moving along a linear periodic track and binding at specific sites located at $x=ld$ ($l=0, \pm1, \pm2, \cdots$) where $d$ is a fixed step distance. In a first treatment of myosin-V on actin filaments we may adopt the observed mean value, $\bar{d} \simeq 36$ nm,  corresponding to to the helix repeat distance (Bray, 2001). However, as discussed further below, the analysis can be extended to take account of the variations in the individual step sizes seen in the data for myosin-V (Mehta {\it et al.}, 1999; Rief {\it et al.}, 2000; Walker {\it et al.}, 2000; Veigel {\it et al.}, 2002): the variation seems primarily to result from binding on actin monomers (at spacing 5.5 nm) adjacent to the main 6.5-monomer helix repeat distance (Steffen {\it et al.}, 2001).

The basic model then supposes that in a catalytic cycle which translocates a motor from binding site $l$ to $l+1$ the protein undergoes $N$ intermediate biochemical transitions from states $j_{l}=0_{l}$ to $1_{l}$ to $2_{l}$ \ $\cdots$ to $(N-1)_{l}$ to $N_{l} \equiv 0_{l+1}$. Kinetic rates $u_{j}$ and $w_{j}$ are associated with the transitions from state $j_{l}$ forwards to state $(j+1)_{l}$ and backwards to state $(j-1)_{l}$, respectively. The state $0_{l}$ represents the motor tightly bound at site $l$ in the absence of fuel molecules --- ATP in the case of myosin-V. Binding of a fuel molecule is represented by the transition $0_{l} \rightarrow 1_{l}$, unbinding by $1_{l} \rightarrow 0_{l}$. Subsequent hydrolysis and release of products occur in the forward transitions $1_{l} \rightarrow 2_{l} \rightarrow \cdots$. But it is important to note that {\it backwards} intermediate transitions {\it and} whole steps (possibly associated with reverse hydrolysis) are allowed {\it and} observed experimentally.

For this model the mean velocity, $V( \{ u_{j}, w_{j} \})$, (see \mbox{Eq. 1}) may be expressed exactly in a closed analytic form in terms of the rate constants $\{ u_{j}, w_{j} \} $ for any value of $N$ ({\bf FK'99}). Furthermore, similar explicit formulae are available for the {\it dispersion} (or effective diffusion constant) of the motion, defined by
\begin{equation}
D=D(\{u_{j},w_{j}\})=\frac{1}{2}\lim_{t \rightarrow \infty} \frac{d}{dt} \left[ \langle x^{2}(t) \rangle - \langle x(t) \rangle^{2} \right].
\end{equation}
This measures the statistical deviation of the motor trajectories from uniform motion at constant velocity. The knowledge of  both the velocity $V$ {\it and} the dispersion $D$, conveniently combined in terms of {\it randomness} (Svoboda {\it et al.}, 1994)
\begin{equation}
r=2D/Vd,
\end{equation}
serves to set bounds on $N$ via a determination of the number of rate-limiting kinetic biochemical transitions and thus yields valuable information regarding the mechanism of processivity (Visscher {\it et al.}, 1999; {\bf KF'00a}; {\bf FK'01}; {\bf FK'02}; Koza, 2002).

To account properly for the externally imposed force, $F$, it is essential ({\bf FK'99}; {\bf FK'01}) to introduce {\it load distribution factors}, $\theta_{j}^{+}$ and $\theta_{j}^{-}$ (for $j=0,1, \cdots, N-1$). Then the transition rates may be taken to vary as
\begin{eqnarray}
u_{j} \Rightarrow u_{j}(F)&=&u_{j}^{0} \exp(-\theta_{j}^{+} Fd/k_{B}T), \nonumber \\
w_{j} \Rightarrow w_{j}(F)&=&w_{j}^{0} \exp(\theta_{j}^{-} Fd/k_{B}T),
\end{eqnarray}
where the most reasonable requirement ({\bf FK'99}; {\bf FK'01}; Hille, 2001) is
\begin{equation}
\sum_{j=0}^{N-1} (\theta_{j}^{+} + \theta_{j}^{-})=1,
\end{equation}
which implies that the condition of stall reflects stochastic quasiequilibrium amongst the (on-pathway) intermediate mechanochemical states.  Indeed, these expressions embody a picture of load-dependent  activation barriers for forward and reverse rates between intermediate states $j$ that lie on a multidimensional reaction pathway. The load distribution factors $\theta_{j}^{\pm}$ provide significant mechanochemical information since they embody a projection of the valleys and cols (or passes) of the reaction pathway onto the force axis, which we suppose is parallel to the motor track. Thus one may identify {\it substeps} of magnitude
\begin{equation}
d_{j}=(\theta_{j}^{+}+\theta_{j+1}^{-})d,
\end{equation}
between motor states $j_{l}$ and $(j+1)_{l}$. If the spatial fluctuations of the center of force of the motor in the intermediate states $j_{l}$ and $(j+1)_{l}$ are sufficiently small relative to $d_{j}$ one may hope to identify this substep in suitably averaged traces $x(t)$ of individual motor motions ({\bf FK'02}).

For the present purpose we note that  the explicit expressions for the mean  velocity, $V$,  for general $N$ lead to a simple relation  for the  stalling  force as defined by $V(F \rightarrow F_{S}) \rightarrow 0$, namely, 
\begin{equation}
F_{S}=\frac{k_{B}T}{d} \ln \left[ \prod_{j=0}^{N-1} (u_{j}^{0}/w_{j}^{0}) \right];
\end{equation}
see {\bf FK'99}.

 The $N$-state periodic kinetic model presented in \mbox{Fig. 1} is, mathematically, an example of the general one-dimensional nearest-neighbor random hopping model for which first-passage questions have been much studied: see van Kampen (1992). Of particular interest here are the so-called splitting probabilities and mean first-passage times. Specifically, in order to analyze observations of motor-protein {\it dwell times}, we need the ``single-step forward splitting probability,'' $\pi_{+}(\{u_{j},w_{j}\})$, defined as the probability that a motor starting at site $l$ will arrive at site $l+1$ {\it without} having undergone sufficiently many intermediate reverse transitions to complete a full backwards step from $l$ to site $l-1$. The corresponding conditional mean single-step first-passage time, $\tau_{+}(\{u_{j},w_{j}\})$, then represents the average time a motor spends at site $l$ before leaving and making a {\it forward} step to site $l+1$. Because of the periodic structure of the $N$-state model of \mbox{Fig. 1} the (rather elaborate) expressions developed by van Kampen (1992) can be simplified considerably even for general $N$ (Kolomeisky and Fisher, to be published). Here we quote the simplest $N=2$ results which will suffice for our present purposes, namely, for the mean forward-step dwell time,
\begin{equation}
\tau_{+}=(u_{0}+u_{1}+w_{0}+w_{1})/(u_{0}u_{1} + w_{0}w_{1}),
\end{equation}
while the fraction of backward (or reverse) steps is
\begin{equation}
\pi_{-}=1-\pi_{+}=w_{0}w_{1}/(u_{0}u_{1} + w_{0}w_{1}).
\end{equation}

Finally we mention that the basic model exhibited in \mbox{Fig. 1} can be extended in various ways while still retaining explicit expressions for $V$, $D$, etc. In particular, one may allow for detachments or ``death'' rates, $\delta_{j}$, from the various motor states and for branching ({\bf KF'00a}), for parallel site-to-site ``jumping'' ({\bf KF'00a}),  for parallel biochemical processes (Kolomeisky, 2001), and for waiting time distributions and  the associated {\it degrees of mechanicity}, $M_{j}^{\pm}$, of the various intermediate processes ({\bf FK'99}; {\bf KF'00b}; {\bf FK'01}). However, the range of observational data so far obtained for myosin-V (unlike that known for kinesin) does not yet warrant consideration of these extensions.

\section*{Analysis of Myosin-V Data}

The bulk-solution  kinetic data on myosin-V ATPase activity indicate that at least two processes, namely, ATP binding and ADP release, should be taken into account in analyzing the   motility (De La Cruz, 1999; Mehta, 2001). While recognizing that a more complete description may require further intermediate states, it is appropriate, therefore to consider first the simplest ($N=2$)-state model. Then, as indicated above, the states $j=0$ correspond to a myosin-V molecule bound to the actin filament in the absence of ATP --- presumably with both heads attached, one behind the other (Walker {\it et al.}, 2000) --- while $j=1$ labels myosin-actin complexes with bound ADP. Thus, in the scheme advanced in Fig. 6 of Mehta (2001), the first and last configurations correspond to $j=0$ while the four intermediate states are gathered into $j=1$; in Fig. 4 of Rief {\it et al.} (2000) the $j=0$ state corresponds to that labelled V; in Fig. 9 of De La Cruz {\it et al.} (2001) the second configuration corresponds to $j=0$, the remaining three to $j=1$.

It now follows that the forward ATP-binding rate should take the form $u_{0}^{0}=k_{0}^{0}$[ATP], where the superscripts 0 denote the limit of zero load: see \mbox{Eq. 5}. On the other hand, the reverse unbinding rate, $w_{1}$, and the forward, ADP release rate, $u_{1}$, should be independent of [ATP], but, of course, may depend on $F$.

According to standard chemical kinetic arguments, the backward rate $w_{0}$ should, in principle, be proportional to [ADP]; and, indeed, the concentration of P$_{\mbox{i}}$ should also play a role. Note, particularly, in this connection the high affinity of ADP for actomyosin which, as discussed by Mehta (2001) and De La Cruz {\it et al.} (2000), had led to significant discrepancies in estimates of steady-state cycling rates. The detailed measurements (Mehta {\it et al.}, 1999; Rief {\it et al.}, 2000) have, therefore, been performed with the aid of an ATP regeneration system [as previously adopted in the kinesin experiments of Visscher et al. (1999)]. In such a set-up neither the concentration of ADP, nor the that of P$_{\mbox{i}}$, is monitored. While experiments that do control [ADP] and [ P$_{\mbox{i}}$] separately are much to be desired, in their absence we are forced (as in {\bf FK'01}) to model the ATP regeneration scheme more or less phenomenologically. Thus if (a) we suppose $w_{0}^{0}=k_{0}'\mbox{[ATP]}^{\alpha}$ (which amounts to [ADP] $\propto$ [ATP]$^{\alpha}$), (b)  recall that the stall force, $F_{S}$, is given by \mbox{Eq. 8}, and (c) note that the current experimental observations reveal no significant dependence of $F_{S}$ on [ATP] (Mehta, 2001), we are led to adopt $\alpha=1$. Indeed, in light of the use of ATP-regeneration  in the experiments, the proportionality of [ADP], and hence of $w_{0}$, to [ATP] at low concentrations is to be expected: see also {\bf FK'01}. It should be remarked, however, that the details of our description of the ATP regeneration scheme play only a minor role in fitting the myosin-V processivity data.

Now in many previous experimental studies of processive motor proteins the mean velocities, $V(\mbox{[ATP]}, F)$, have been measured and reported. Such observations must, at least in principle, include some fraction of backward or reverse steps, especially at large loads approaching stall. However, in their experiments on myosin-V (Mehta {\it et al.}, 1999; Rief {\it et al.}, 2000) the authors opted to measure only dwell times, separating adjacent steps of mean size $d \simeq 36$ nm (Mehta, 2001), {\it preceding forward steps}. Thus their reported dwell times, $\tau(\mbox{[ATP]}, F)$, as plotted in \mbox{Fig. 2A}, do {\it not} precisely correspond to an ``overall mean step time,'' say $\bar{\tau}$, related to the mean velocity simply via $\bar{\tau} = d/V$ --- although at low loads, where the fraction of reverse steps is small, $\bar{\tau}$ should provide a good approximation; but under near stall conditions, when $V \rightarrow 0$, the overall mean step time, $\bar{\tau}$, diverges to infinity whereas the dwell times $\tau( F$$\rightarrow$$ F_{S})$ remain bounded. Rather, we identify the observed dwell times with the conditional single-step mean first-passage times, $\tau_{+}$, identified above: see \mbox{Eq. 9}. Accordingly, our analysis of the myosin-V data is based upon the expression
\begin{equation}
\tau(F,\mbox[ATP])= \frac{k_{0}^{0} \mbox{[ATP]}e^{-\theta_{0}^{+} Fd/k_{B}T}+u_{1}^{0}e^{-\theta_{1}^{+} Fd/k_{B}T} + k_{0}'\mbox{[ATP]}e^{\theta_{0}^{-} Fd/k_{B}T}+w_{1}^{0}e^{\theta_{1}^{-} Fd/k_{B}T}}{k_{0}^{0} \mbox{[ATP]} e^{-\theta_{0}^{+} Fd/k_{B}T}  u_{1}^{0} e^{-\theta_{1}^{+} Fd/k_{B}T} + k_{0}'\mbox{[ATP]}e^{\theta_{0}^{-} Fd/k_{B}T} w_{1}^{0}e^{\theta_{1}^{-} Fd/k_{B}T}},
\end{equation}
following from Eqs. 5, 6, and 9, with $d=\bar{d}=36$ nm. 

Then, by systematically exploring the full seven-dimensional parameter space specified by ($k_{0}^{0}, \cdots, \theta_{0}^{-}$) we find that the observed stall force, $F_{S}$, and the dynamics of myosin-V as a function of [ATP] and of the load, $F$, up to $F_{S}$,  are well described by the rates
\begin{eqnarray}
k_{0}^{0}=0.70 \pm 0.10 \ \mu\mbox{M}^{-1}\mbox{s}^{-1}, &  u_{1}^{0}=12.0 \pm 1.0 \mbox{ s}^{-1}, \\ \nonumber
k_{0}'= (5.0 \pm 0.5) \times 10^{-6}  \ \mu\mbox{M}^{-1}\mbox{s}^{-1}, &  \quad w_{1}^{0}= (6.0 \pm 0.5) \times 10^{-6} \mbox{ s}^{-1}, 
\end{eqnarray}
and the load-distribution factors
\begin{eqnarray}
\theta_{0}^{+}=-0.010 \pm 0.010, & \quad \theta_{1}^{+}=0.045 \pm 0.010, \\ \nonumber 
\theta_{0}^{-}= 0.580 \pm 0.010, & \quad \theta_{1}^{-}=0.385 \pm 0.010.
\end{eqnarray}
It should be noted that consideration of the limits of low and high [ATP] and low and high loads, confirm a fair degree of independence of the various fitting parameters. The uncertainties indicated in Eqs. 12 and 13 correspond to the ranges of acceptable fits to the processivity data while constraining the other parameters appropriately. The central values yield the fits presented  in \mbox{Fig. 2} as solid curves. 

In respect to our fits for $k_{0}^{0}$ and $u_{1}^{0}$ note that the bulk solution kinetic experiments yield an ATP binding rate constant (corresponding to $k_{0}^{0}$) between 0.7 and 1.6 $\mu$M$^{-1}$s$^{-1}$, while the ADP-release rate (corresponding to $u_{1}^{0}$) is about 12-16 s$^{-1}$ (Mehta, 2001; De La Cruz {\it et al.}, 1999). The agreement is clearly most satisfactory.

\section*{Discussion}
\subsection*{Mean Velocity and Load Dependence}

The quality of the fits in \mbox{Fig. 2} ensures that the observed (approximate) Michaelis-Menten behavior is respected. Indeed, using the rate and load-distribution parameters in Eqs. 12 and 13 and previous theory (e.g., {\bf FK'01}) enables us to predict the variation of the mean velocity, $V$, with $F$ and [ATP]: see the solid curves \mbox{Fig. 3}. Evidently, the  stall force of about 3 pN seen in the experiments is reproduced. Note also, from the dwell-time data imposed on the predictions in \mbox{Fig. 3} using $V \simeq d/ \tau$, that, as anticipated in the discussion before Eq. 11, the approximation $\tau \simeq \bar{\tau} \equiv d/V$ is valid for small loads (up to $F \simeq 2.5$ pN). Indeed, from Eq. 10  (with Eqs. 5, 6, 12, and 13) one finds that the fraction of reverse steps is negligible until $F \gtrsim 2.5$ pN. 

\subsection*{Load Dependence of Rates}

It is notable from \mbox{Eq. 13} that within the fitting uncertainties there is essentially {\it no} load-dependence to the binding of ATP to the myosin-V-actin complex, i.e., $\theta_{0}^{+} \simeq 0$; see also Mehta (2001). This contrasts strongly with the properties of conventional kinesin moving on a microtubule where $\theta_{0}^{+} \simeq 0.13$ was found in {\bf FK'01} for both $N=2$ and $N=4$ fits. This lack of load-dependence on binding ATP to actin-myosin accounts for the fact that the dwell time remains constant at saturating ATP conditions up to $F \simeq 2.3$ pN: see Fig. 2{\it A}.

Nevertheless the other transitions {\it are} load dependent with ADP release bearing a modest ($ \sim 5 \%$) fraction of the dependence. In parallel to kinesin, however --- see {\bf FK'01}, the reverse transitions carry most of the load-dependence. Indeed, the load distribution pattern ({\bf FK'01}) for myosin-V is close to a featureless descending ramp. Note that this result is in striking contrast to the implications of the phenomenological expression Eq.\ 1 which suggests that only forward (i.e., binding and/or hydrolysis) processes need be considered and could  exhibit significant load-dependence. Indeed, our analysis indicates that at least three biochemical transitions in the actin-myosin-V ATPase cycle are load-dependent whereas Eq.\ 1 entails only a single load-dependent process.  It seems that this difference is the main reason why fits for the ``characteristic distance'' $d'$ in  Eq.\ 1 differ so markedly from the true mean step size $\bar{d} \simeq 36$ nm. Since our analysis recognizes reverse transitions, which, by the fits, occur at a non-vanishing rate that is  enhanced under load [see, again, Eq. 10], an explanation is provided for the observation of more frequent backward steps in myosin-V at high loads (Rief et al., 2000). Our treatment also provides a basis for a quantitative discussion of the ADP inhibition effect which it would be instructive to explore further experimentally.

\subsection*{Substeps}

A striking feature of the data of Rief {\it et al.} (2000) is the observation of ``half steps'' under high loads ($\gtrsim 2$ pN). From the published traces the steps appear to correspond to an intermediate state with a mean center of force lying a distance, say $d_{1/2}$ forward from the bound-state ($j=0$) center with $d_{1/2}/d \simeq 0.48 \pm 0.04$. On the other hand, \mbox{Eq. 7} and the load distribution factors in \mbox{Eq. 13} indicate a substep with $d_{0}/d \simeq 0.38 \pm 0.03$ (corresponding to $d_{0} \simeq 13-14$ nm). Rief {\it et al.} (2000) suggest that these ``half steps'' (always followed by a complementary forward or backward step to complete a movement with $\langle \Delta x \rangle = d$ or 0) reflect an ``off-pathway state'' because they remain rare even under the high loads that uncover their presence. While this suggestion seems most reasonable on the available evidence, our analysis suggests that the half steps might possibly represent genuine substeps (lying on or close to the main reaction pathway), which appear stochastically under high loads when  the forward rates, $u_{1}(F)$, have been  slowed down while the reverse rates $w_{1}(F)$ are significantly enhanced.

In other experiments Veigel {\it et al.} (2002) observed attachments of single myosin-V molecules to an actin filament (stretched between two optically trapped beads) at [ATP]=100 $\mu$M. After some of the  attachment events, ``staircases'' of from two or three to a dozen forward steps were seen of mean size 36 nm; the staircases typically terminated in an effective stall (signaled by interspersed forward and backward steps) before detachment from the filament: see Fig.\ 3 of Veigel {\it et al}. However, the authors  concluded that the {\it first step} in each staircase had a mean size of only $d_{1}=26.2 \pm 2.3$ nm (similar to the amplitude of isolated attachment events lacking any subsequent steps). A similar initial unitary step of $\sim 20$ nm was seen in experiments by Moore {\it et al.} (2001) on heavy meromyosin-like fragments of myosin-V. These displacements were identified (in both articles) as a ``working stroke''; and Veigel {\it et al.} saw a comparable step of $\sim 21$ nm in attachment events of a {\it single-}headed recombinant myosin-V. Furthermore, Veigel {\it et al.} in their Fig.\ 5A, report stiffness measurements (using a sinusoidal driving force) which revealed low-stiffness intervals of variable durations (longer at higher loads): the mid-positions of these intervals was about 20 nm further along the actin filament than the preceding higher-stiffness intervals, a displacement similar to the initial ``working stroke.''

In our formulation and fits using a single intermediate mechanochemical state prior to completion of a full ($d=36$ nm) step, such a $d_{1}$ should as the notation chosen suggests, correspond to a $d-d_{0} \simeq 22$ nm ``substep.'' The agreement of these  various findings (within the combined experimental and fitting uncertainties) appears to lend support to our values for the load-distribution factors $\theta_{j}^{\pm}$. However, corresponding substeps have not been identified at  low loads by Mehta, Rief and coworkers.  Nevertheless, a detailed examination of the sample stepping records for [ATP] = 2 mM and $F$ = 1 pN presented in Fig. 2A of Rief \emph{et al.} (2000) reveals plausible indications of substeps in 13 to 16 of a total of around 32 ``full" steps of $\sim36$ nm, some of the substeps appearing to have dwell times as long as 0.1 - 0.2 s.  More favorable   conditions for detecting the predicted substeps and checking their dwell times should be realized  at low loads and [ATP] $\simeq 10 \mu$M (which corresponds roughly to the effective Michaelis-Menten concentration, $K_M$: see Rief \emph{et al.} (2000) and Fig. 3). Such data not consistent with the  present predictions might require the introduction of waiting-time distributions (\textbf{KF'00b}): see also the remarks below concerning randomness.

\subsection*{Variability of Step Sizes}

The fits to the data so far described have utilized a fixed step size, $d$, taken equal to the observed mean step size $\bar{d} \simeq 36$ nm that corresponds closely, as mentioned above, to the known (half) repeat distance of the actin filament double helix (Bray, 2001). But separate single-molecule experiments by Steffen et al. (2001) using myosin-S1 motor domains indicate ``target zones'' for binding to the filament consisting of three adjacent accessible actin monomers at spacings $\Delta d \simeq 5.5$ nm, the active zones repeating along the filament helices at $\sim$36 nm intervals. Furthermore, the processivity data for myosin-V reveal significant variations in individual step sizes about the mean, $\bar{d}$. The observations [see: Mehta {\it et al.} (1999) Table 1; Rief {\it et al.} (2000) \mbox{Fig. 2B}; Walker {\it et al.} (2000) \mbox{Fig. 2}; Veigel {\it et al.} (2002) \mbox{Fig. 3b}] are consistent with about 60$\%$ of the steps being of size $d_{(0)}=36$ nm while 20$\%$ each are of sizes $d_{(\pm)}=d_{(0)} \pm \Delta d = 41.5$ and 30.5 nm; only a few percent of longer or shorter steps appear. The fact that 40-45$\%$ of the observed  steps deviate from $d_{(0)}=36$ nm raises the possibility that our fits using a unitary step size might be misleading or especially sensitive to the spread in sizes.

To address this issue note, first, that steps of distinct sizes, say $d_{(k)}$, should be expected to have different mean dwell times: an ideal set of experimental observations would, then, report the corresponding $\tau_{(k)}(F,\mbox{[ATP]})$ and their probabilities, say $p_{k}$. An analysis using Eq. 11 with $d$ replaced by $d_{(k)}$, etc., could subsequently be performed for each set and might possibly prove revealing. To a leading approximation one may suppose the various dwell times will be independent: in that case, the overall mean dwell time should be given by
\begin{equation}
\tau= \sum_{k} p_{k} \tau_{(k)}.
\end{equation}

More realistically, however, if the target-zone picture is valid, there will be correlations between successive steps: thus on average a short step, say of size $d_{(-)}$, must be followed immediately by a longer step, of size $d_{(0)}$ or $d_{(+)}$, and vice-versa. In principle, such correlations are open to observation and one might, indeed, expect the dwell times to depend on the size of the {\it previous step}, say $d_{(k)}'$, as well as on the step to be made. Theoretically the situation could clearly be modeled by a Markov process. [See, e.g., Steffen et al. (2001).]

In the absence of such more detailed observations, however, we may test the sensitivity of our fits by further exploratory calculations. As an extreme case, suppose 50$\%$ of the steps are of magnitude $d_{(+)}=41.5$ nm and 50$\%$ of size $d_{(-)}=30.5$ nm. How would the fits change from those assuming a unitary step $d_{(0)}=36$ nm? An answer is displayed by the dashed curves plotted in Fig. 2. These have been obtained by using Eq. 14 with $p_{+}=p_{-}=1/2$ and computing $\tau_{(+)}$ and  $\tau_{(-)}$ from Eq. 11 using $d_{(+)}$ and $d_{(-)}$ together with the {\it same} zero-load rates and load distributions factors given in Eqs. 12 and 13. As evident from Fig. 2, there is no significant change in the quality of the fits --- even though it would not be unreasonable to suppose that the rates and load factors might have some dependence on the $\pm 15\%$ changes in step-size. One might say that ``the averages win out'' ---  a not unexpected conclusion.

In fact we may go further and study the effects of {\it correlated step sizes} by utilizing the expressions for $N$-state periodic models ({\bf KF'03}) with $N$ an integral multiple of $N_{0}$, the number of intermediate states in the basic catalytic cycle. In our analysis we have $N_{0}=2$ and so can utilize an $N=2+2=4$ periodic system to describe {\it alternating long and short steps} of sizes $d_{(+)}$ and $d_{(-)}$ (with, of course, the same previous {\it average} step size $d_{(0)}$). If we again use the zero-load rates and distribution factors in Eqs. 12 and 13, and compute the mean velocity as a function of load, we obtain the dashed curves presented in Fig. 3. Once more the deviations from the $d=\bar{d}$ results are  negligible at loads $F < 2$ pN, while at higher loads sufficiently precise data might reveal discrepancies.

We conclude, therefore, that the consequences of replacing a distribution of step sizes by the mean $\bar{d}$ are not significant at current levels of experimental precision. Conversely, unless fairly  precise experimental data can be obtained  that are categorized by step length, there may be little more that can be reliably determined by  fitting such observations.

\subsection*{Randomness}

As mentioned previously, the fluctuation statistics of  motor motion are  effectively captured in the randomness parameter, $r$, as defined in \mbox{Eq. 4}. The fits presented in \mbox{Eqs. 12} and 13 suffice to predict the variation of $r$ with [ATP] under various loads (or vice versa) {\it assuming} that all the rate processes may be adequately represented as standard kinetic transitions: see {\bf FK'99} and {\bf KF'00}. The corresponding predictions for $r(\mbox{[ATP]})$ are  presented in \mbox{Fig. 4} for loads $F=0.4$ and 2.5 pN. At low [ATP] the randomness is  close to unity indicating   that only one  rate-limiting  process is effective in this concentration range. However, under a low load a marked dip to $r \simeq 0.5$ occurs around $ \mbox{[ATP]}=10-20 \ \mu$M: this, in turn, is indicative of two competing rate processes that both play a role in this ``crossover'' regime. On the other hand, at high loads that approach stall, $r$ rises rapidly above unity; however, this is primarily a consequence of the vanishing of the velocity $V$ when $F \rightarrow F_{S}$ since $r$ must then diverge: see also Fig. 5(B), below.

It must be noted, however, that the analogous  predictions, on the basis of an ($N=2$)-state kinetic model, for the  the randomness of kinesin are {\it not} supported by the data of Visscher {\it et al.} (1999). Rather, for low loads and $\mbox{[ATP]} \gtrsim 30 \ \mu$M, the randomness falls rapidly and {\it remains below} 0.5 up to saturation concentrations: because of the bound $ r \ge 1/N$ ({\bf FK'99}, Koza, 2002), this is inconsistent with a kinetic description. Thus the data for  conventional kinesin demand $N=4$ (or more) states (in accord with the usual biochemical picture of ATP hydrolysis). {\it Alternatively}, and, in light of certain experiments (Nishiyama {\it et al}., 2001), possibly more realistically, one may invoke a waiting-time distribution to describe the process of hydrolysis and ADP release with a {\it mechanicity} $M_{1} \simeq 0.6$ ({\bf KF'00}, {\bf FK'01}). Thus measurements of $r(F,\mbox{[ATP]})$ for myosin-V might well prove equally revealing of the mechanism by {\it failing} to verify the behavior predicted by \mbox{Fig. 4}!

\subsection*{Reverse or Assisting Loads}

Another interesting and potentially instructive set of predictions can be advanced for the behavior under {\it negative} or {\it assisting} loads, $F<0$. Such experiments have been performed for kinesin by Coppin et al. (1997). Although their data posed certain problems (in particular, a significantly lower overall processivity under low loads) the same load distribution factors (and similar rates) provided a not unreasonable ($N=2$) fit ({\bf FK'01}) simply by extending the analog of \mbox{Eq. 11} to negative values of $F$. The corresponding predictions for the dwell time and for the randomness as a function of $F$, extending down to $-3$ pN, are displayed in \mbox{Fig. 5}.

A caveat must, however, again be raised in light of subsequent experiments on kinesin by Block (2001) and coworkers. The validity of the extension of \mbox{Eq. 11} to negative $F$ clearly rests on a mechanistic/geometric assumption, namely, that changing abruptly the direction at which the coiled-coil myosin tail leaves the junction with two heads (or motor domains), i.e., from trailing upwards and backward  ($F>0$) to pulling upwards and forward ($F<0$) does not result in a corresponding abrupt change in the mechanics of ATP binding, unbinding, or hydrolysis. If the junction were a perfect universal swivel joint, then as $F$, which is just the {\it component} of the total load force, say $ \vec{F}$, parallel to the track, passes through zero the stresses and strains within motor should, indeed, vary smoothly. However, the junction cannot be totally torsion free and if, for example, the tail were to rest against part of the head in one configuration but become dissociated in the other, then the smoothness assumption embodied in \mbox{Eqs. 5} would fail. Indeed, just such an abrupt change of behavior has since been found by Block and coworkers (2001) for kinesin. Clearly, comparable experiments on myosin-V are desirable and should prove informative.

\section*{Conclusions}

In summary, we have presented a simple two-state stochastic model, with allowance for  fluctuating step sizes, which describes well essentially all the available experimental data on single-molecule myosin-V processivity. It reveals that ATP binding is load-independent, while ADP release is weakly load-dependent; but (as for kinesin) the loading forces  strongly  affect the reverse transition rates. Our analysis  is consistent with the observation of tight coupling between catalytic cycles and mechanical steps, i.e., one ATP molecule is consumed per individual  step, and with ATP binding and ADP release rates measured in bulk solution. It also indicates  that an  intermediate myosin-ADP-actin complex has its center of force advanced by  13-14 nm forward from the position prior to ATP binding, in reasonable agreement with various observations indicating  a subsequent ``working stroke'' of around 22 nm. We have discussed  specific predictions for the dwell times, mean velocity, and randomness  of myosin-V motors  in various  experimental regimes including the imposition  of assisting loads.  Further experiments are needed in order to investigate  the validity of our theoretical description and to uncover other mechanochemical features of myosin-V.

\section*{Acknowledgments}

Discussions with Jonathon Howard and Steven Block concerning kinesin and insightful comments from John Sleep and the referees concerning myosin-V have been appreciated. A.B.K.  acknowledges the financial support of the Camille and Henry Dreyfus New Faculty Awards Program (under Grant NF-00-056). M.E.F. is grateful to the National Science Foundation for support (under Grant No. CHE-99-81772).

\newpage

\hspace{5cm}{\bf FIGURE LEGENDS}
\vspace{1cm}

\noindent \mbox{FIGURE  1}. \hspace{0.4em} Specification of the simplest $N$-state periodic stochastic model. A motor  in state $j_{l}$ can undertake a forward transition at  rate $u_{j}$ or it can make  a  backward transition at rate $w_{j}$. The bound state $N_{l}$ is identified with $0_{l+1}$.

\vspace{5mm}

\noindent \mbox{FIGURE  2}. \hspace{0.4em}  Fits to the data of Mehta {\it et al.} (1999) for the  mean dwell times of myosin-V: ({\it A})  as a function of external load, $F$, at different ATP concentrations; ({\it B})  as a function of [ATP] under an external load $F=0.4$ pN and a  prediction for $F=2.3$ pN. The solid curves represent Eq. 11 with the central parameter values in Eqs. 12 and 13; the dashed curves represent the mean dwell times predicted for a 50:50 mixture of short, $d_{(-)}=30.5$ nm, and long, $d_{(+)}=41.5$ nm steps using the same values for the other parameters: see the subsection {\it Variability of Step Sizes}, below. (Note that in part ({\it B}) the dashed curve for $F=0.4$ pN cannot be distinguished from the solid curve.)
\vspace{5mm}

\noindent \mbox{FIGURE  3}. \hspace{0.4em} The force-velocity or ($F$, $V$) dependence of myosin-V at various concentrations of ATP as predicted using  the parameter values in Eqs. 12 and 13: solid curves.  The corresponding dashed curves follow from  a model with {\it alternating} long and short steps ($d_{(+)}=41.5$ nm and $d_{(-)}=30.5$ nm) but otherwise the same zero-load rate constants and load distribution factors, $\theta_{j}^{\pm}$. The superimposed data bars (for [ATP]$=1 \mu$M and 2 mM) derive from the observed dwell times by using the {\it approximate} relation $V \simeq d/\tau$ (with $d=36$ nm); they track the predictions for $V(F)$ fairly well because of the paucity of reverse or backward steps under loads $F \lesssim 2.5$ pN.
\vspace{5mm}

\noindent \mbox{FIGURE  4}. \hspace{0.4em} Predictions for the variation of the randomness, $r$, of myosin-V as a function of [ATP] at low ($F=0.4$ pN) and high external load ($F=2.3$ pN).
\vspace{5mm}

\noindent \mbox{FIGURE  5}. \hspace{0.4em} Predicted  behavior of myosin-V under assisting (i.e., negative) and resisting (positive) external loads, $F$,  for two ATP concentrations: ({\it A}) mean dwell time; ({\it B})  randomness. See the text for appropriate caveats.

\begin{figure}[t]
\begin{center}
\vskip 1.5in
\unitlength 1in
\begin{picture}(4.5,3.3)
\resizebox{4.5in}{3.3in}{\includegraphics{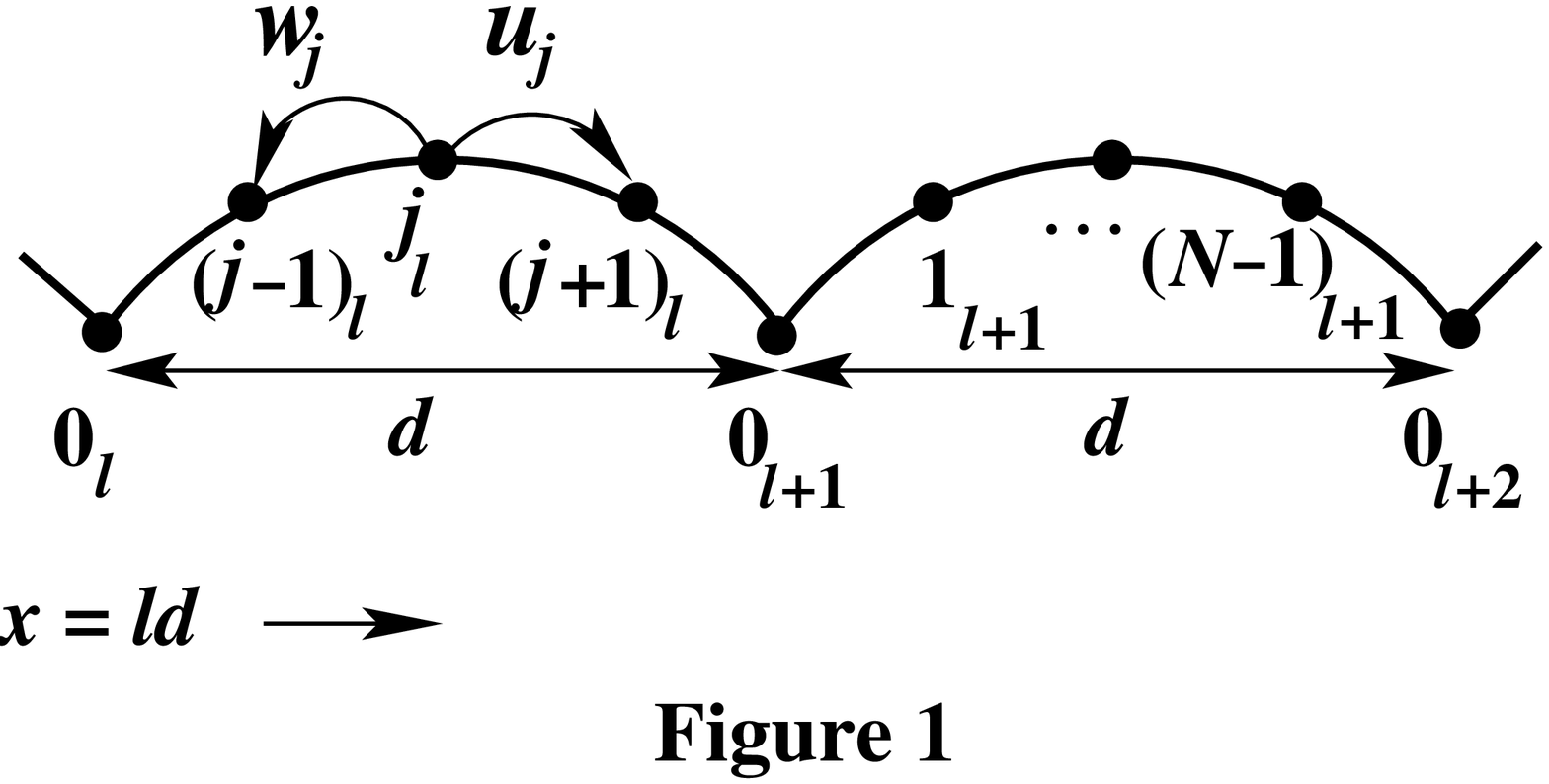}}
\end{picture}
\vskip 3in
 \begin{Large} Fig.1   \end{Large}
\end{center}
\vskip 3in
\end{figure}

\begin{figure}[t]
\begin{center}
\vskip 1.5in
\unitlength 1in
\begin{picture}(4.5,3.3)
\resizebox{4.5in}{3.3in}{\includegraphics{Fig2A.myosin.eps}}
\end{picture}
\vskip 3in
 \begin{Large} Fig.2A   \end{Large}
\end{center}
\vskip 3in
\end{figure}

\begin{figure}[t]
\begin{center}
\vskip 1.5in
\unitlength 1in
\begin{picture}(4.5,3.3)
\resizebox{4.5in}{3.3in}{\includegraphics{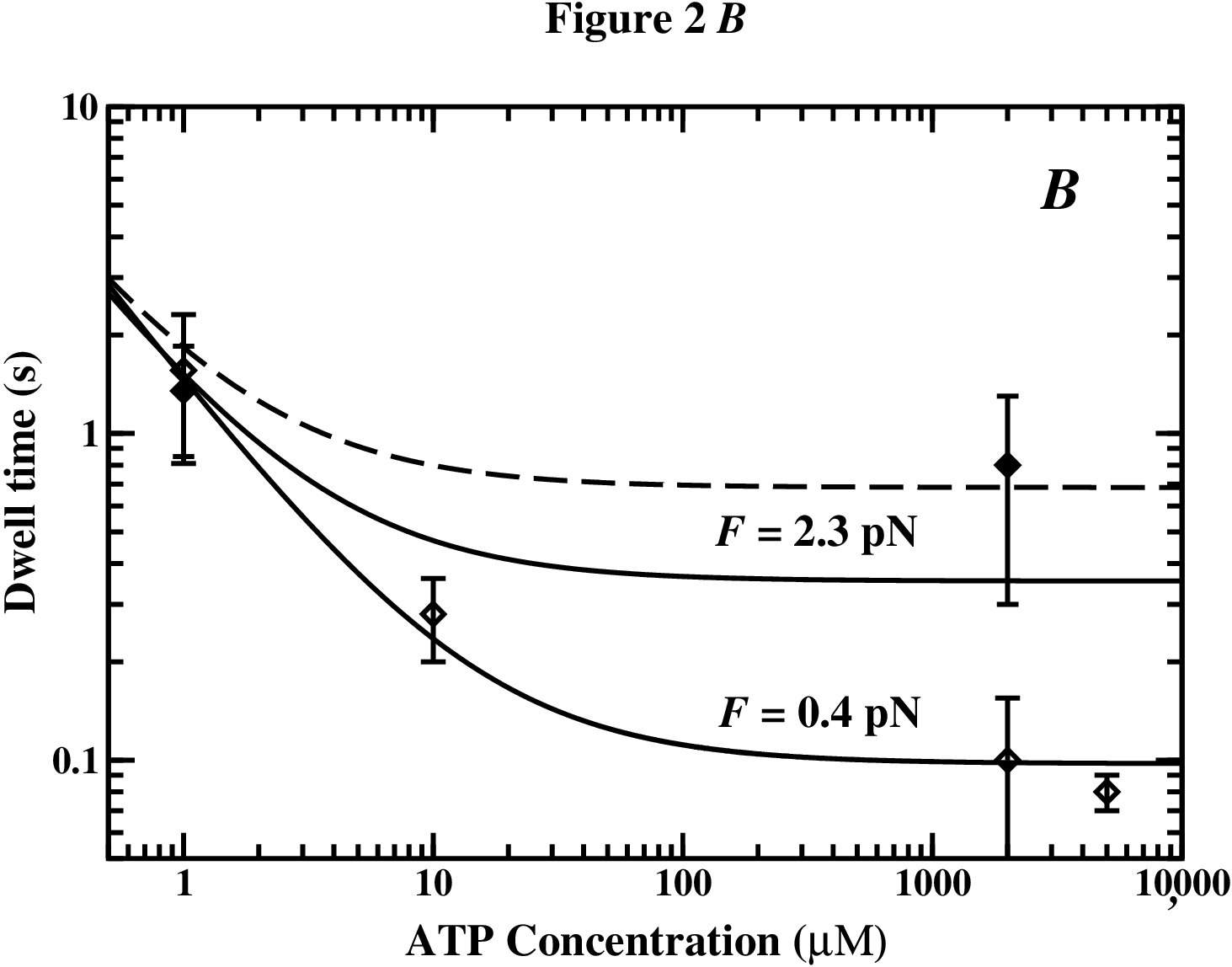}}
\end{picture}
\vskip 3in
 \begin{Large} Fig.2B  \end{Large}
\end{center}
\vskip 3in
\end{figure}

\begin{figure}[h]
\begin{center}
\vskip 1.5in
\unitlength 1in
\begin{picture}(4.5,3.3)
\resizebox{4.5in}{3.3in}{\includegraphics{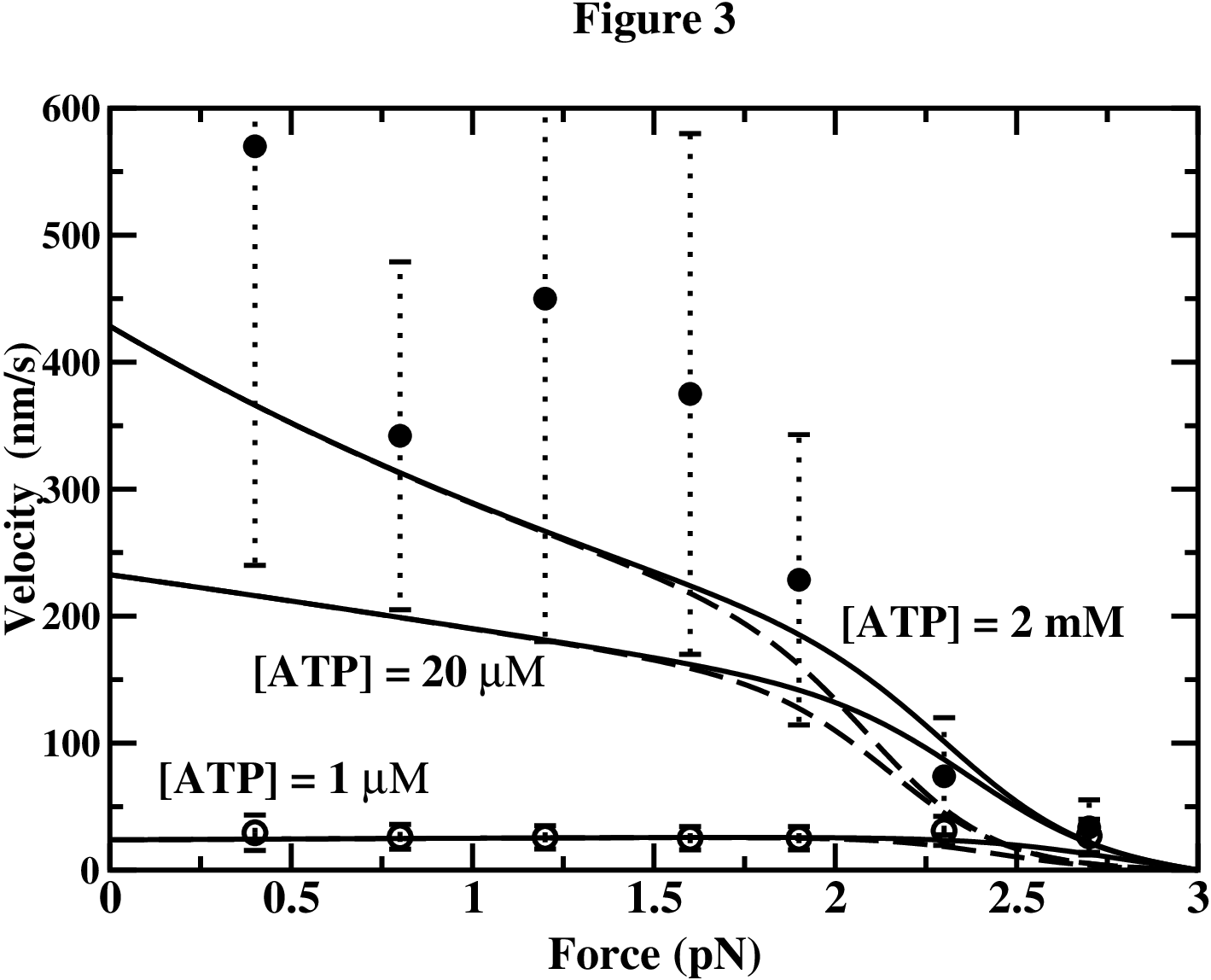}}
\end{picture}
\vskip 3in
 \begin{Large} Fig.3  \end{Large}
\end{center}
\vskip 3in
\end{figure}

\begin{figure}[t]
\begin{center}
\vskip 1.5in
\unitlength 1in
\begin{picture}(4.5,3.3)
\resizebox{4.5in}{3.3in}{\includegraphics{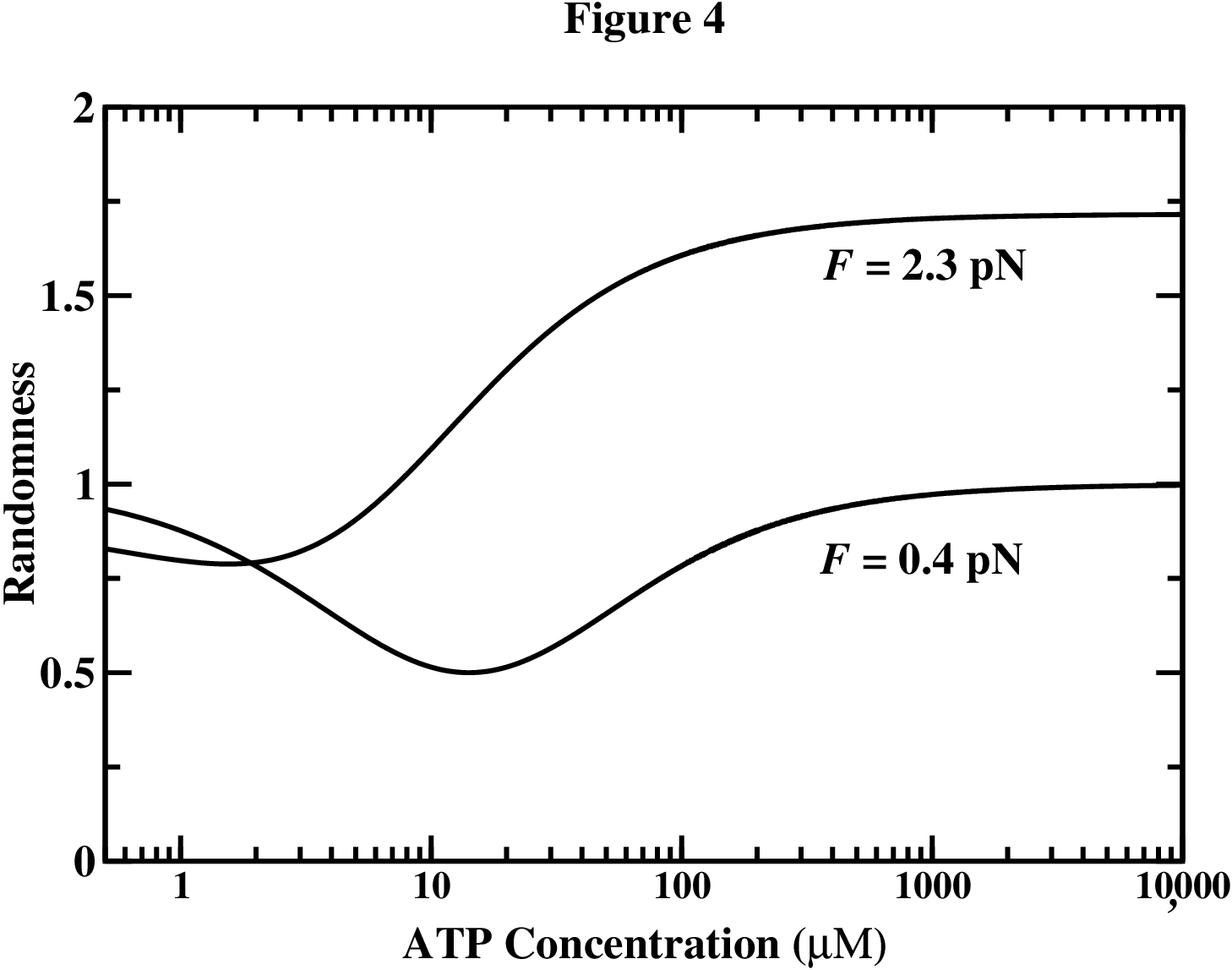}}
\end{picture}
\vskip 3in
 \begin{Large} Fig.4  \end{Large}
\end{center}
\vskip 3in
\end{figure}

\begin{figure}[t]
\begin{center}
\vskip 1.5in
\unitlength 1in
\begin{picture}(4.5,3.3)
\resizebox{4.5in}{3.3in}{\includegraphics{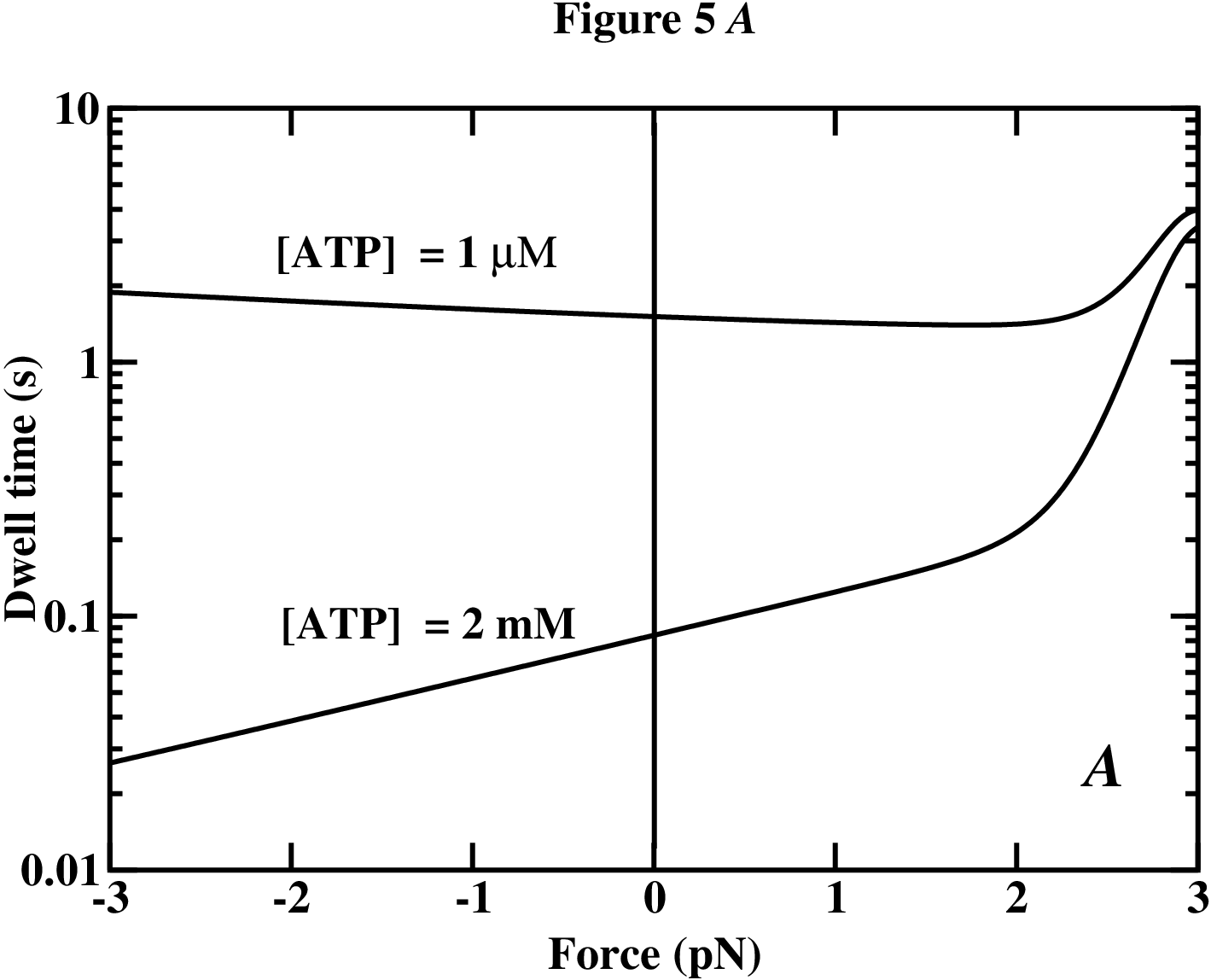}}
\end{picture}
\vskip 3in
 \begin{Large} Fig.5A  \end{Large}
\end{center}
\vskip 3in
\end{figure}

\begin{figure}[t]
\begin{center}
\vskip 1.5in
\unitlength 1in
\begin{picture}(4.5,3.3)
\resizebox{4.5in}{3.3in}{\includegraphics{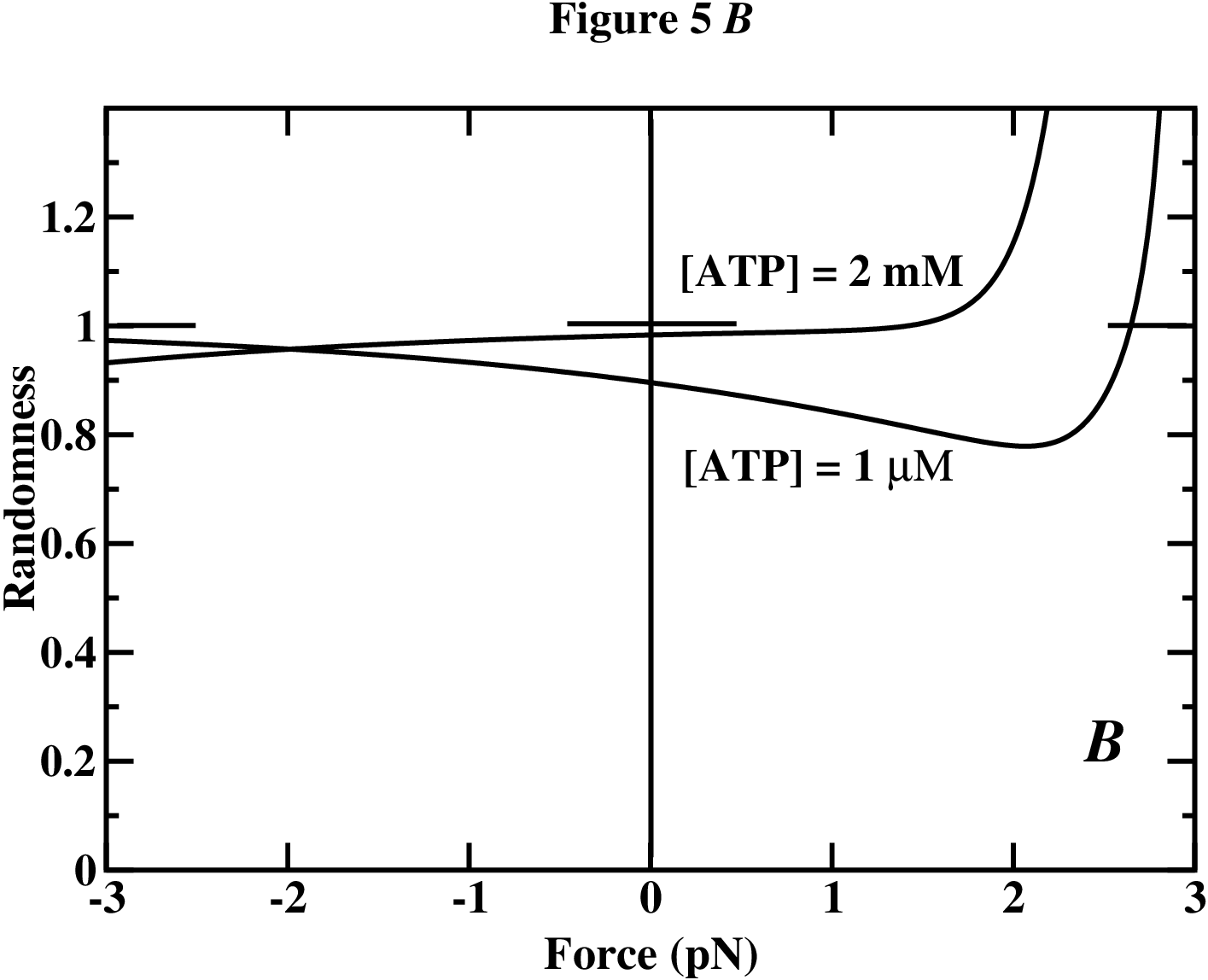}}
\end{picture}
\vskip 3in
 \begin{Large} Fig.5B \end{Large}
\end{center}
\vskip 3in
\end{figure}

\end{document}